\documentclass[12pt]{iopart}
\usepackage{epsf}
\usepackage{graphicx}
\begin{document}

\title[A pulsed source of continuous variable polarization entanglement]{A
pulsed source of continuous variable polarization entanglement}

\author{Oliver Gl\"ockl\dag\, Joel Heersink\dag\, Natalia
Korolkova\dag\, Gerd Leuchs\dag\, Stefan
Lorenz\dag\ \footnote[3]{stefan.lorenz@physik.uni-erlangen.de}}

\address{\dag\ Zentrum f\"ur Moderne Optik, Physikalisches Institut,
Universit\"at Erlangen--N\"urnberg, Staudtstra{\ss}e 7/B2, 91058 Erlangen, Germany}

\begin{abstract}
We have experimentally demonstrated polarization entanglement using
continuous variables in an ultra-short pulsed laser system at the telecommunications wavelength of 1.5~$\mu$m.
Exploiting the Kerr non-linearity of a glass fibre we generated a polarization squeezed pulse with $\hat{S}_2$
as the only non-zero Stokes parameter thus $\hat{S}_1$ and $\hat{S}_3$ being the conjugate pair.
Polarization entanglement was generated by interference of the polarization squeezed field with a
vacuum on a 50:50 beam splitter. The two resultant beams exhibit strong quantum noise correlations
in $\hat{S}_1$ and $\hat{S}_3$. The sum noise signal of $\hat{S}_3$ was at the respective shot
noise level and the difference noise signal of $\hat{S}_1$ fell -2.9~dB below this value.
\end{abstract}

\maketitle

\section{Introduction}
The emerging field of quantum information processing relies on certain
quantum mechanical state properties. Particularly
entanglement of two or more sub states cannot be described classically and is
crucial for quantum information and communcation protocols such as
teleportation and cryptography. The first experimental realizations of entangled
states used two two-level systems (qubits) \cite{Wu50}. In recent years, not
only these discrete systems were improved, but also a new type emerged:
continuous variable systems \cite{OuPer92}. These use continuous quantum
observables such as amplitude and phase quadratures of the electromagnetic
field for the entanglement in analogy to position and momentum of the original
EPR gedankenexperiment \cite{EinPod35}.\\ In the case of intense light fields
the polarization can be described by a set of continuous variables which can be
entangled. The advantage of polarization over quadrature entanglement is the
ease of detection, which does not require a phase reference such as a local
oscillator. All relevant polarization parameters can be determined by passive
setups using direct detection \cite{KorLeu01a}.\\ The first experiments
on polarization squeezing used continuous wave light and parametric processes
\cite{HalSor01, Bow02b}. In silica fibres quadrature
squeezing \cite{RosShe91,BerHau91,KorLeu00,Sch98} and polarization squeezing
\cite{Her03} have been shown experimentally. Entanglement can be generated from squeezing using
passive elements such as beam splitters. Entanglement of quadratures has also
been achieved \cite{OuPer92,SilLam01}. Polarization entanglement of continuous
wave light was recently shown by Bowen et al. \cite{Bow02c} by transformation of
quadrature entanglement. In this paper we present a source for pulsed
polarization entanglement which is compact and stable.

\section{Polarization entanglement}
To describe the quantum polarization state of an intense light field, one can
use the quantum Stokes operators \cite{JauRoh55, Rob74, ChiOrl93, UsaSod01},
which are derived from the classical Stokes parameters \cite{Sto52}. If
$\hat{a}_{x/y}$ and $\hat{a}^{\dagger}_{x/y}$ denote the photon annihilation and
creation operators of two orthogonal polarization modes x and y, and
$\hat{n}_x$ and $\hat{n}_y$ are the photon number operators of these modes, the
quantum Stokes operators read as follows: \begin{eqnarray}
\hat{S}_{0} &=& \hat{a}^{\dagger}_{x} \hat{a}_{x} + \hat{a}^{\dagger}_{y}
\hat{a}_{y} = \hat{n}_x + \hat{n}_y, \nonumber \\
\hat{S}_{1} &=& \hat{a}^{\dagger}_{x} \hat{a}_{x} - \hat{a}^{\dagger}_{y}
\hat{a}_{y} = \hat{n}_x - \hat{n}_y, \nonumber \\
\hat{S}_{2} &=& \hat{a}^{\dagger}_{x} \hat{a}_{y} + \hat{a}^{\dagger}_{y}
\hat{a}_{x}, \nonumber \\
\hat{S}_{3} &=& i(\hat{a}^{\dagger}_{y} \hat{a}_{x} - \hat{a}^{\dagger}_{x}
\hat{a}_{y}).
\end{eqnarray}
The operators $\hat{S}_1$, $\hat{S}_2$, and
$\hat{S}_3$ follow the operator valued commutation relation of a SU(2) Lie
algebra:
\begin{equation}
\left[ \hat{S}_k,\hat{S}_l\right]  = 2i\varepsilon_{klm}\hat{S}_m.
\end{equation}
This gives rise to a set of three Heisenberg-type uncertainty relations:
\begin{eqnarray}   V_{1} V_{2} \geq
|\langle\hat{S}_3\rangle|^{2}, \qquad V_{3} V_{1} \geq
|\langle\hat{S}_2\rangle|^{2},  \qquad   V_{2} V_{3} \geq
|\langle\hat{S}_1\rangle|^{2}.   \label{eqn:stokes_uncertainty}
\end{eqnarray}
Thus, for a light beam with non--zero $\langle\hat{S}_2\rangle$, as in our
experiment, the values of $\hat{S}_1$ and
$\hat{S}_3$ cannot be determined with arbitrary
accuracy. The variance $V_j=\langle\hat{S}_j^2\rangle -
\langle\hat{S}_j\rangle^2$ of $\hat{S}_j$ cannot vanish for $j=1$ and $j=3$
simultaneously. A state which obeys
\begin{equation}
V_k < |\langle\hat{S}_l\rangle| < V_m, \qquad k\neq l \neq m
\end{equation}
is a polarization squeezed state (\cite{KorLeu01a} and references therein).\\
Polarization entanglement of two intense light fields can be characterized in
two ways, both derived from the characterization of quadrature entanglement. One possibility is to check if one can infer the value of a
noncommuting observable of one subsystem from a measurement on the other
subsystem of the pair to a precision better than given by the Heisenberg
uncertainty relation \cite{Rei89}. This is called EPR entanglement. The measure
of the precision of such inference is
the conditional variance $V_{cond}$ for the subsystems A~and~B
\begin{equation}
V_{cond}\left(\hat{S}_{k,A} \vert  \hat{S}_{k,B}\right)=\left\langle {\left(
\delta \hat{S}_{k,A}\right) }^2\right\rangle - \frac{{\vert \langle \delta
\hat{S}_{k,A} \delta \hat{S}_{k,B} \rangle \vert}^2}{\langle {( \delta
\hat{S}_{k,B}) }^2\rangle},
\end{equation}
where we defined
\begin{equation} \hat{S}_k =
\langle\hat{S}_k\rangle + \delta \hat{S}_k. \end{equation}
A state
is then EPR entangled if \cite{KorLeu01a,Bow02c}:
\begin{equation}
V_{cond}(\hat{S}_{1,A}|\hat{S}_{1,B})
V_{cond}(\hat{S}_{3,A}|\hat{S}_{3,B})<
{\vert \langle \hat{S}_{2,B} \rangle \vert}^2. \label{EPR}
\end{equation}
The other entanglement criterion was derived by Duan \cite{DuaGie00a} and
Simon \cite{Sim00} as an extension of the Peres-Horodecki
non--separability criterion for continuous variables.
A state with nonvanishing $\langle\hat{S}_2\rangle$ is called
polarization entangled if
\begin{equation}
V\left(\hat{S}_{1,A}-\hat{S}_{1,B}\right)+V\left(\hat{S}_{3,A}+\hat{S}_{3,B}
\right ) < 2 \vert \left\langle \hat{S}_{2,A}\right\rangle \vert + 2 \vert
\left\langle \hat{S}_{2,B}\right\rangle \vert. \label{PeresHorodecki}
\end{equation}
A state which is non--separable according to equation
(\ref{PeresHorodecki}) can be generated by the interference of a polarization
squeezed light field with a vacuum field on a 50:50 beam splitter (see figure
\ref{polentsetup}). If the polarization squeezed beam is composed of two
such orthogonally polarized amplitude squeezed beams the resulting beam is
polarized along the $S_2$ direction. These beams exhibit quantum correlations
in $\hat{S}_1$ and $\hat{S}_3$. As the input vacuum has an uncertainty identical
to that of a coherent state (i.e. it is not squeezed), the variance
$V\left(\hat{S}_{3,A}+\hat{S}_{3,B}\right)$ corresponds
to the shot noise of a coherent beam. However the variance
$V\left(\hat{S}_{1,A}-\hat{S}_{1,B}\right)$ drops below this shot
noise level. If we were to feed an intense coherent field or a
polarization squeezed light field intstead of a vacuum into the second input
port of the beam splitter, then $V\left(\hat{S}_{3,A}+\hat{S}_{3,B}\right)$
would also show non--classical correlations. The advantage of using a vacuum
input is experimental ease, since using an bright beam implies the necessity of
a further phase lock but also produces improved correlations.

\section{Experimental setup}
\begin{figure}[h] \begin{center} \epsfxsize=10cm
\epsfbox{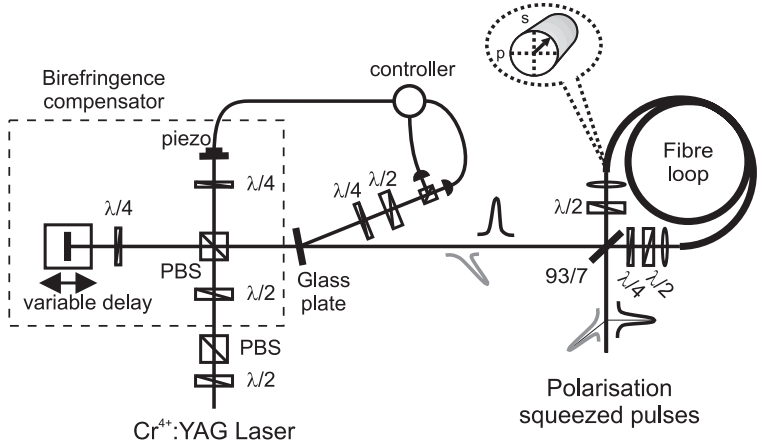}
\end{center}
\caption{\label{polsqueezingsetup}Birefringent compensator and nonlinear fiber Sagnac interferometer.}
\end{figure}

\begin{figure}[h]
\begin{center}
\epsfxsize=14cm
\epsfbox{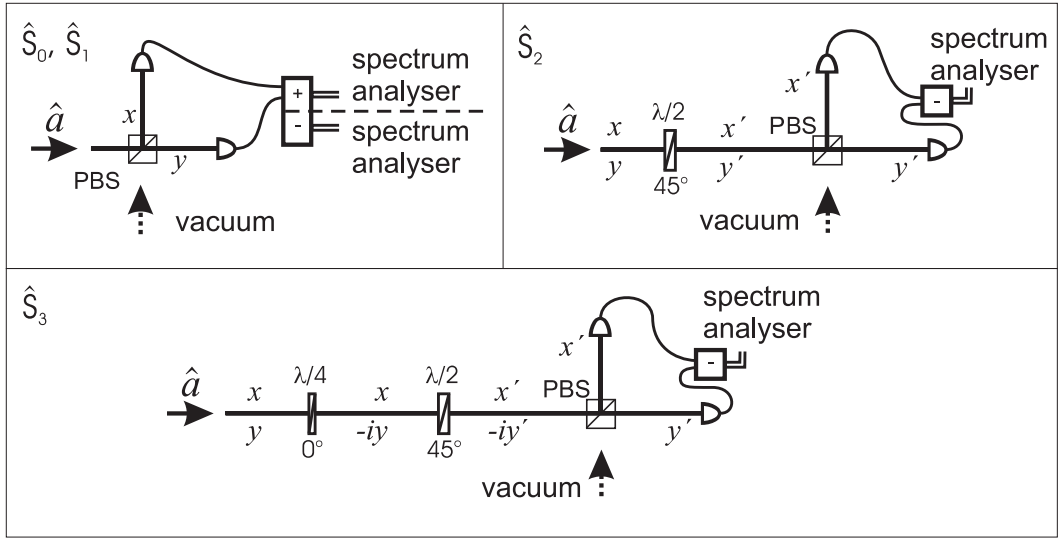}
\end{center}
\caption{\label{Smessung}Detection setups for the Stokes parameters. The unknown
polarization state $\hat{a}$ is split at a polarizing beam splitter and measured
on two detectors. Upper left: $\hat{S}_0$ and $\hat{S}_1$; upper right:
$\hat{S}_2$; below: $\hat{S}_3$.}
\end{figure}

\begin{figure}[h] \begin{center} \epsfxsize=9cm
\epsfbox{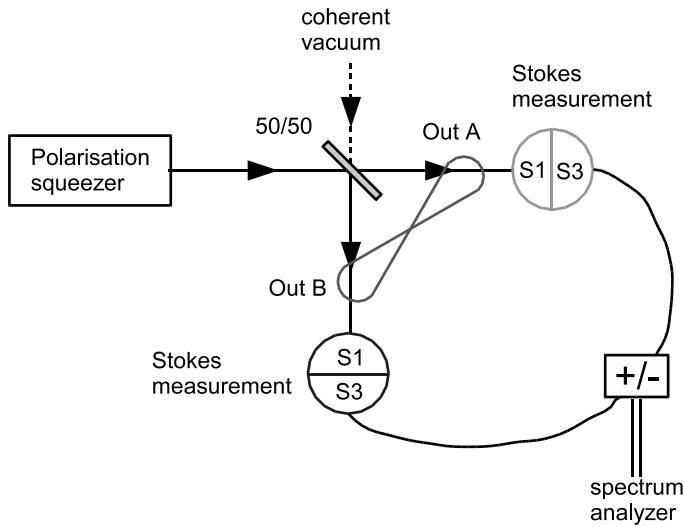} \end{center}
\caption{\label{polentsetup}Setup for the generation of polarization
entanglement. In the two output ports, A~and~B, the Stokes parameters $S_1$ and
$S_3$ were measured. The photocurrents were added/subtracted to check for
correlations.} \end{figure}

In the experiment a Cr:YAG laser with a wavelength 1495~nm was used. It produced
soliton shaped pulses (163~MHz) with a duration of 150~fs by passive modelocking
through a semiconductor saturable absorber \cite{SpaBoh97}. Those pulses are
coherent and thus shot noise limited.
To produce polarization squeezing two amplitude squeezed beams are required.
These are generated in an asymmetric fibre Sagnac interferometer
\cite{Sch98}, which consist of a 93:7~beam splitter and 14.2~meters of
fibre (3M-FS-PM-7811). The use of a polarization maintaining fibre allows the
generation of two independent amplitude squeezed light fields
of orthogonal polarization \cite{KorLeu00,SilLam01}. However, due to the fibre birefringence
the emerging pulses are temporally separated. Thus a polarization
dependent delay (birefringence compensator) was inserted before the
interferometer. As the two beams emerge from the same fibre, an excellent
spatial overlap is achieved. The temporal coincidence is actively controlled by
measuring the reflected light at the interferometer input and
correspondingly adjusting the birefringence compensator. If the phase of the two
emerging beams is synchronized, a polarization squeezed beam
with polarization in $S_2$ direction is produced \cite{Her03} (see figure
\ref{polsqueezingsetup}).\\
This polarization squeezed beam is mixed
with vacuum on a 50:50~beam splitter. The two resulting intense beams are
directed into independent Stokes measurement setups labelled A~and~B (see figure
\ref{polentsetup}). Each consists of two identical detectors, a
polarization beam splitter and two optional retardation elements
($\frac{\lambda}{2},\frac{\lambda}{4}$) to measure the
fluctuations of $\hat{S}_1$ or $\hat{S}_3$ respectively \cite{KorLeu01a} (see
figure \ref{Smessung}).
The detected AC photocurrents are passively added or subtracted and
monitored on two spectrum analyzers (HP 8590E, measurement frequency 17.5~MHz,
300~kHz resolution bandwidth, 30~Hz video bandwidth). To determine the degree of
polarization squeezing the 50:50~beam splitter was removed.

\section{Results}

\begin{figure}[h]\label{polsq}
\begin{center}
\epsfxsize=14cm
\epsfbox{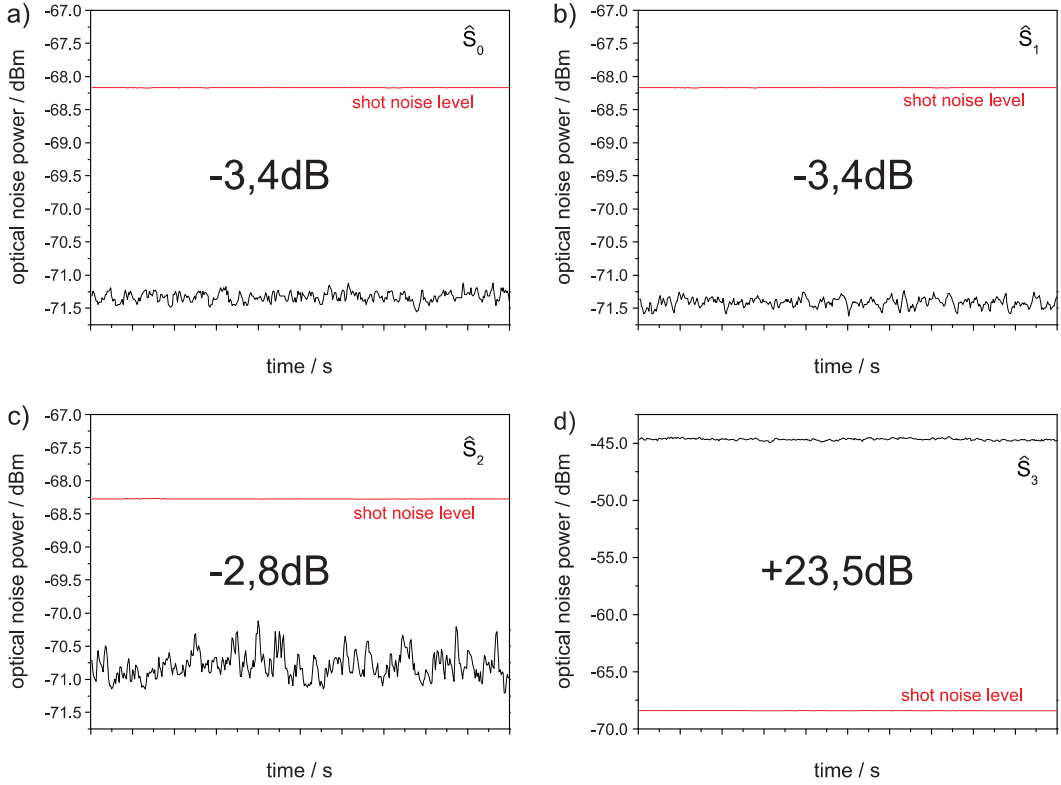}
\end{center}
\caption{Characterization of the polarization squeezing source. $\hat{S}_0$ and
$\hat{S}_2$ are amplitude squeezed, $\hat{S}_1$ is polarization squeezed and
$\hat{S}_3$ is anti--squeezed.}
\end{figure}
\begin{figure}[h]\label{s1}
\begin{center}
\epsfxsize=14cm
\epsfbox{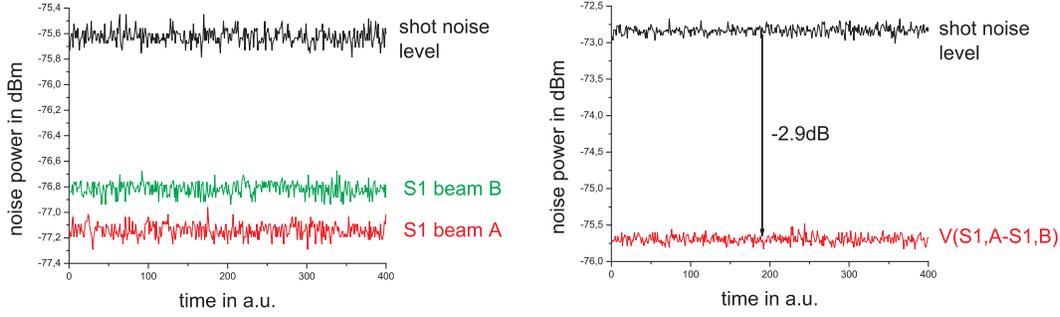}
\end{center}
\caption{Polarization squeezing (left) and correlations (right) in the
$\hat{S}_1$ parameter. The difference noise $V(S_{1,A}-S_{1,B})$ is 2.9~dB below
shot noise.}
\end{figure}

Polarization squeezing of -3.4~dB
in the $\hat{S}_1$ parameter was observed, while its canonic conjugate, the
$\hat{S}_3$ parameter, is anti--squeezed by +23.5~dB (see figure \ref{polsq}).
The noise traces to characterize polarization squeezing as well as those for the
polarization entanglement were corrected for electronic noise which was
-86.9~dBm.\\ Polarization entanglement was generated using the scheme described
above. As the specific polarization squeezed state from our setup has a nonzero
$\hat{S}_2$ mean value, equation (\ref{PeresHorodecki}) can be used to check
for non--separability. The non--classical correlations in the conjugate
Stokes parameters were observed by measuring the respective Stokes parameters
at the two output ports of the beam splitter and taking the variance of the sum
and the difference signals. In figure \ref{s1} the variance of
$\hat{S}_{1,A}-\hat{S}_{1,B}$ is plotted as well as the variances of the Stokes
parameters of the individual modes at the output ports A and B and the
corresponding shot noise level. The difference signal is 2.9~dB below the shot
noise level. Each individual mode is already squeezed in $\hat{S}_1$, but only
by -1.3~dB due to the contribution of the vacuum fluctuations. Non--classical
correlations in the $\hat{S}_1$ parameter are observed and we found
$V(\hat{S}_{1,A}-\hat{S}_{1,B})=0.52$. Note that the noise traces of the
polarization squeezing and the polarization entanglement experiments can not
be compared directly, as additional electronic rf--splitters/combiners, which
attenuate the detected photocurrents, are necessary in the polarization
entanglement setup.

The noise traces of the $\hat{S}_3$ parameter are rather different. Each
individual signal at the two output ports has a high degree of noise, as the
initial beam was anti--squeezed in the $\hat{S}_3$ parameter. Nevertheless, the
variance of the sum signal $\hat{S}_{3,A}+\hat{S}_{3,B}$ coincides with
the shot noise level. Thus, the squeezing variance
$V(\hat{S}_{3,A}+\hat{S}_{3,B})=1$.
\begin{figure}[h]
\begin{center}
\epsfxsize=14cm
\epsfbox{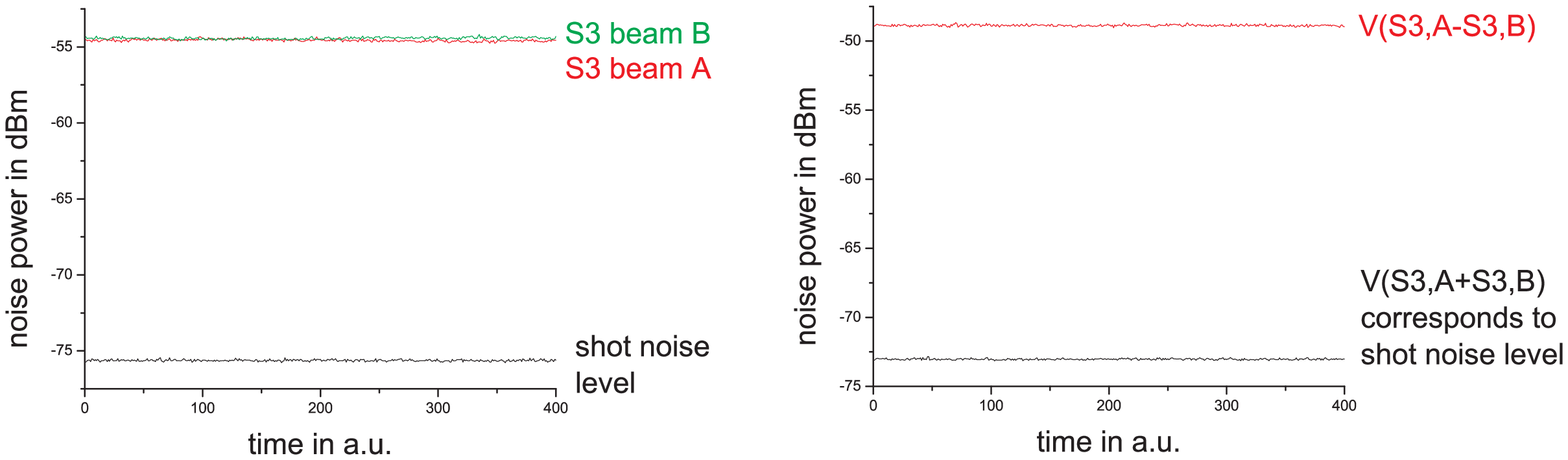}
\end{center}
\caption{Polarization anti--squeezing (left) and correlations (right) in the
$\hat{S}_3$ parameter. The sum noise $V(S_{3,A}+S_{3,B})$ is at the shot noise
level.} \end{figure}
The applicaton of the non--separability criterion of equation
(\ref{PeresHorodecki}), \begin{equation}
\frac{V(\hat{S}_{1,A}-\hat{S}_{1,B})+V(\hat{S}_{3,A}+\hat{S}_{3,B})}{ \langle
\hat{S}_{2,A}\rangle + \langle \hat{S}_{2,B} \rangle } = 0.52 + 1 < 2,
\label{entanglementgemessen} \end{equation}
proves that a highly correlated
non--separable quantum state in the Stokes variables has been generated.

\section{Conclusions and outlook}
We have shown that the source described above produces two intense light fields
which are entangled in their polarization variables. Thus the entanglement was
detected and manipulated without the need of a stable phase reference as it is
the case for quadrature entanglement. All relevant parameters were
checked in direct detection. In contrast to sources using, e.g. optical
parametric amplifiers, only one phase has to be locked, to achieve good 
stability. Only one nonlinear device (a polarization maintaining fibre) is
needed to produce the entanglement, making the source compact.\\ The degree of
entanglement can be further improved in two ways. To generate sub shot
noise quantum correlations in both conjugate variables one needs to combine the
polarization squeezed beam with a bright coherent beam instead of the vacuum.
Both variances in equation (\ref{entanglementgemessen}) would then drop below 1
as it is desireable for many applications of polarization entanglement in
quantum communication protocols. The complexity of the experiment would increase
only moderately, as one additional phase would be locked. To increase the degree
of polarization entanglement even further, the interference of two polarization
squeezed beams at the 50:50~beam splitter is necessary. This would result in a
polarization entanglement equal to the degree of the amplitude squeezing
invested, in our case more than 3~dB. The price one has to pay is the need for
another birefringent compensator and fibre Sagnac interferometer.\\ The source
is especially suited for future quantum communication experiments as it produces
entangled states deterministically and at a high repetition rate limited only by
the laser repetition rate. A further fibre integration of the source would
simplify its introduction into existing communication networks. Thus its use in
further quantum information and communication experiments for teleportation and
quantum cryptography is very promising.

\ack
This work was supported by the Schwerpunkt Programm 1078 of the
Deutsche Forschungsgemeinschaft and by the EU grant under QIPC, project
IST-1999-13071 (QUICOV). We thank T. Gaber for extensive experimental help with
the polarization squeezing setup and Ch. Marquardt for technical assistance.

\section*{References}

\bibliographystyle{unsrt}

\end{document}